\documentclass[preprint,review,3p,12pt]{elsarticle}

\usepackage{amsmath}
\usepackage{graphics}

\journal{Chaos, Solitons and Fractals}

\begin{document}

\begin{frontmatter}

\title{On the dynamics of bubbles in boiling water}

\author[uem,inct]{H.V. Ribeiro}
\ead{hvr@dfi.uem.br}
\author[uem,inct]{R.S. Mendes}
\author[uem,inct]{E.K. Lenzi}
\author[uem]{M.P. Belancon}
\author[uem]{L.C. Malacarne}
\address[uem]{Departamento de F\'isica, Universidade Estadual de Maring\'a, Av. Colombo 5790, 87020-900, Maring\'a, PR, Brazil}
\address[inct]{National Institute of Science and Technology for Complex Systems, CNPq, Rua Xavier Sigaud 150, 22290-180, Rio de Janeiro, RJ, Brazil}

\begin{abstract}
We investigate the dynamics of many interacting bubbles in boiling water by using a
laser scattering experiment. Specifically,  we analyze the temporal variations of a laser intensity signal 
which passed through a sample of boiling water. Our empirical results indicate that the return interval
distribution of the laser signal does not follow an exponential distribution; contrariwise, a heavy-tailed distribution has been found. 
Additionally, we compare the experimental results with those obtained from a minimalist phenomenological
model, finding a good agreement.
\end{abstract}

\end{frontmatter}

\section{Introduction}
Bubbles are ubiquitous in nature and their dynamics is both
fascinating and very complex\cite{Prosperetti,Lohse}. It is not surprising that 
bubbles are an effervescent source of research.
For instance, bubbles appear in the context of energy generation\cite{Joshi,Alvarez},
collapsing bubbles can emit ligth\cite{Brenner}, turbulent thermal
convection has been observed in a single soap bubble\cite{Seychelles}, cooperative and avalanches-like dynamics
are present in collapsing of aqueous foams\cite{Vandewalle, Ritacco},  and singularities emerge
when air bubbles detache from a nozzle submerged in water\cite{Schmidt}.

Despite the fact that the fundamental equations ruling the behavior of moving
fluids are well known, an analytical or even a numerical approach can become 
infeasible for many common situations. This is particularly true for
bubbles in turbulent fluids, where, at higher Reynolds numbers, the number
of mesh points required to solve each bubble as well as the flow around it grows
up leading to a long simulation time\cite{Tryggvason}.

A very familiar case, where bubbles appears in such contexts, is the boiling process
of water\cite{Zahn}. For exemple,
when the temperature reaches 100$^\circ$C the vapor pressure is 1 bar and we can observe
the spontaneous process of bubble formation (nucleation)\cite{BrennerC}. 
Although it is an ordinary process, nucleate boiling has several complex aspects involving thermal
interactions between bubbles and the heated surface and among the nucleation sites. There are also
hydrodynamics interactions bubble to bubble and bubble to liquid bulk\cite{Shoji}. 
In this scenario, simple models, whether phenomenological or not, and simple experiments have been
designed to try to clarify this intricate dynamics. For instance, extensive studies have been done 
by considering nonlinear models and experiments of boiling\cite{Shoji} as well as evaporation in microchannels\cite{Thome}
or in short capillary tube\cite{Cordonet}. However,
as far the authors know, much less attention has been paid to clouds of bubbles, i.e.,
a system containing many  interacting bubbles (see for instance Ref.\cite{Iida}).
Our main goal here is attempt to fill this hiatus by using a simple experiment that basically 
consists of a laser beam passing through a sample of boiling water. We also comfront
the experimental results with a minimalist model towards improving our understanding of this complex system.

This article is organized as follows. Primarily, we describe the experimental setup and the
data acquisition. Next, we report a statistical analysis of the data and present a model.
Finally, we end this work with some concluding comments and a summary.

\section{Experimental setup and data presentation}
\begin{figure*}[ht]
\centering
\includegraphics[scale=0.45]{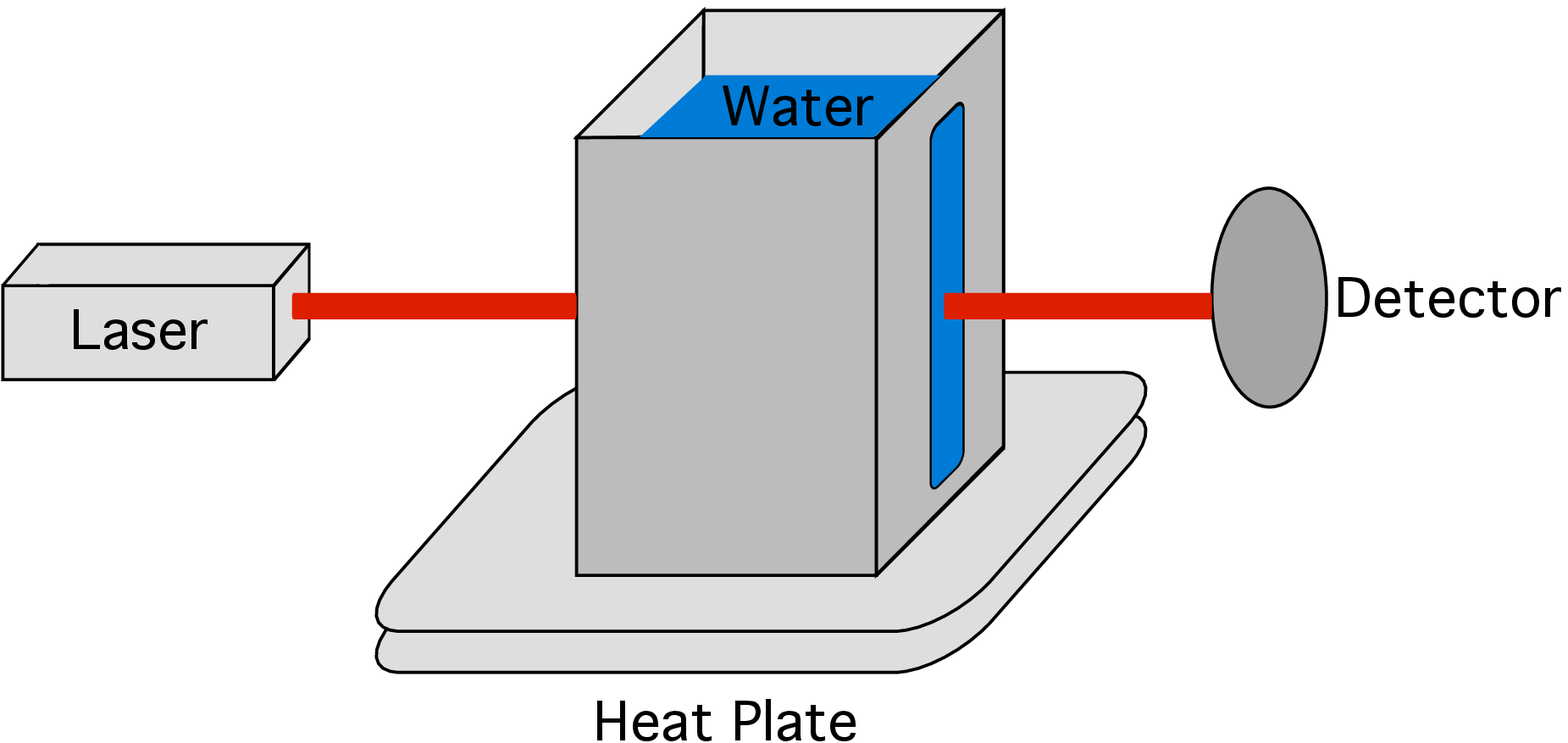}
\includegraphics[scale=0.4]{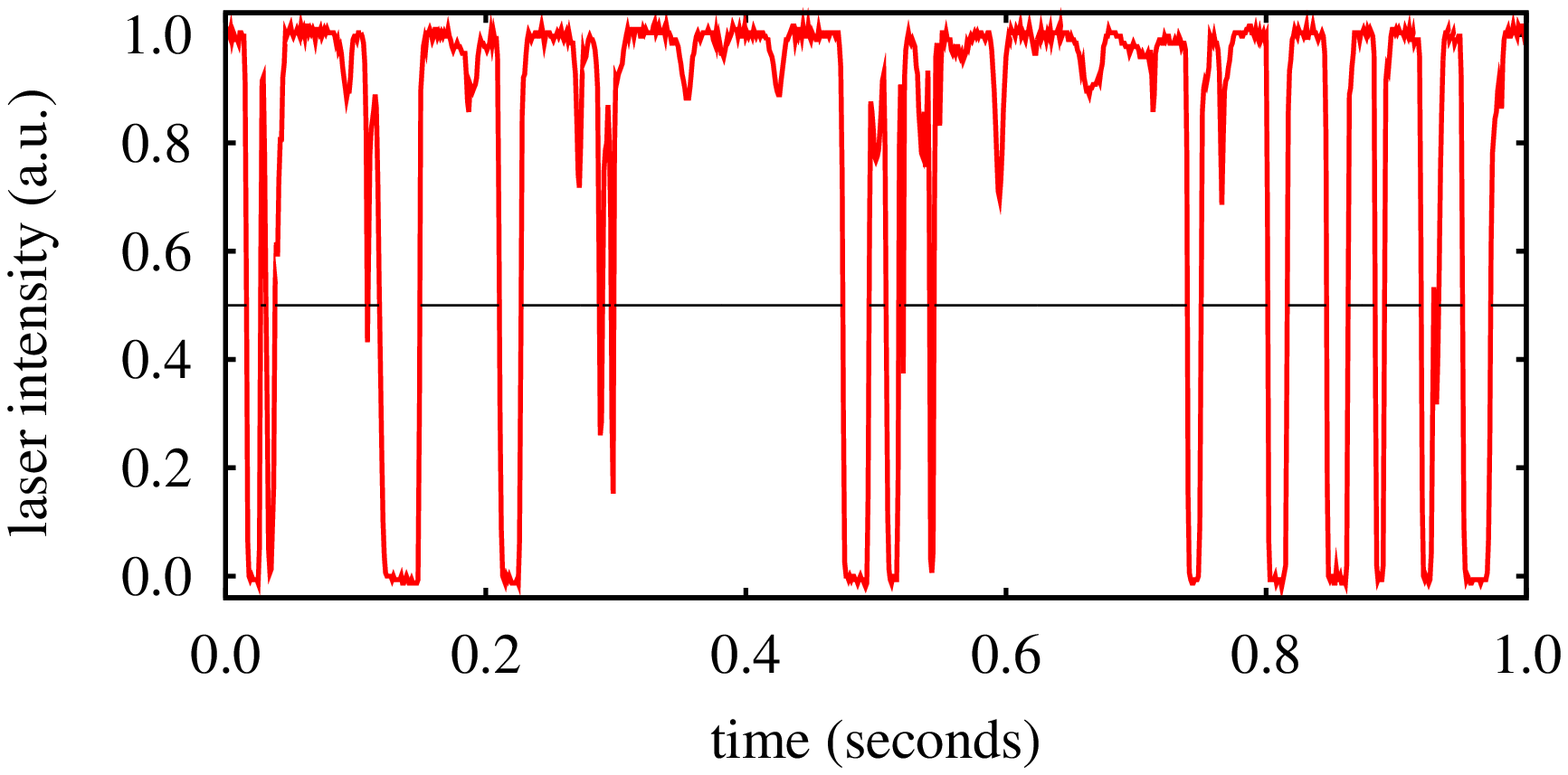}
\caption{(left panel) A schematic representation of the experimental setup: 
	a laser beam passes through a sample of boiling water which is in contact with the heat plate.
	The laser intensity is monitored by the detector (photodiode) and recorded. (right panel) A
	typical laser intensity signal for a sample of approximately 300 ml of boiling water.}\label{fig:exp}
\end{figure*}

The experiment consists of samples of approximately 300 ml of distilled water at atmospheric pressure and confined
by impermeable metallic walls with glass windows through which the laser beam is transmitted, as shown in Figure \ref{fig:exp} (left panel). 
At the bottom, the confining vessel is in contact with a heat plate that employs a power of around 300 W, in such
way that the temperature in the contact interface is approximately 300 $^\circ$C. After the boiling
process becomes stable, i.e., the water temperature stabilizes, we start to record the He-Ne laser
(10 mW) intensity signal that passes through the sample by using a photodiode detector (Thorlabs DET100A) coupled to an oscilloscope
(Tektronix TDS5032B) with a sampling rate of a thousand points per second. Typical recording times are of the order of 10 minutes
and the height of the incident beam does not significantly change the statistical results, avoiding the water-air interface.

Figure \ref{fig:exp} (right panel) shows a typical record signal. We can see that the signal is characterized by intermittent valleys.
{ While the emerging dynamic is complex, the individual processes 
generating the bubble are qualitatively simple. At the button of the confining vessel the temperature is higher what makes a
small fraction of the liquid to evaporate, producing the bubbles. These nucleation sites are not static and depend on the heat 
transfer and also on the liquid-wall interactions. The  bubbles depart from the nucleation sites and rise through the confining vessel.
Along this movement the bubbles continuously interact with each other, with the liquid and with the walls. When one or more
bubbles crossing the laser path they scatter the light, producing a decrease in the record signal.}
Naturally, this intrincated signal reflects the complex collective dynamical behavior of the bubbles.
Similar situations are customary when dealing with time series. For instance, the earth seismic and geomagnetic activity  can be investigated by considering a seismogram and the DST index.

\section{Data analysis}
Due the conditions of the signal, a natural variable to investigate the system dynamics is the time difference between
the extreme events characterized by sharped valleys. This analysis is frequently employed in the physics and financial
literature\cite{Gumbel,Galambos,Reiss,Embrechts} and it shows to be useful when investigating the underling mechanism 
ruling the system\cite{Bunde,Yamasaki,Wang,Blender}.

\begin{figure*}[!ht]
\centering
\includegraphics[scale=0.42]{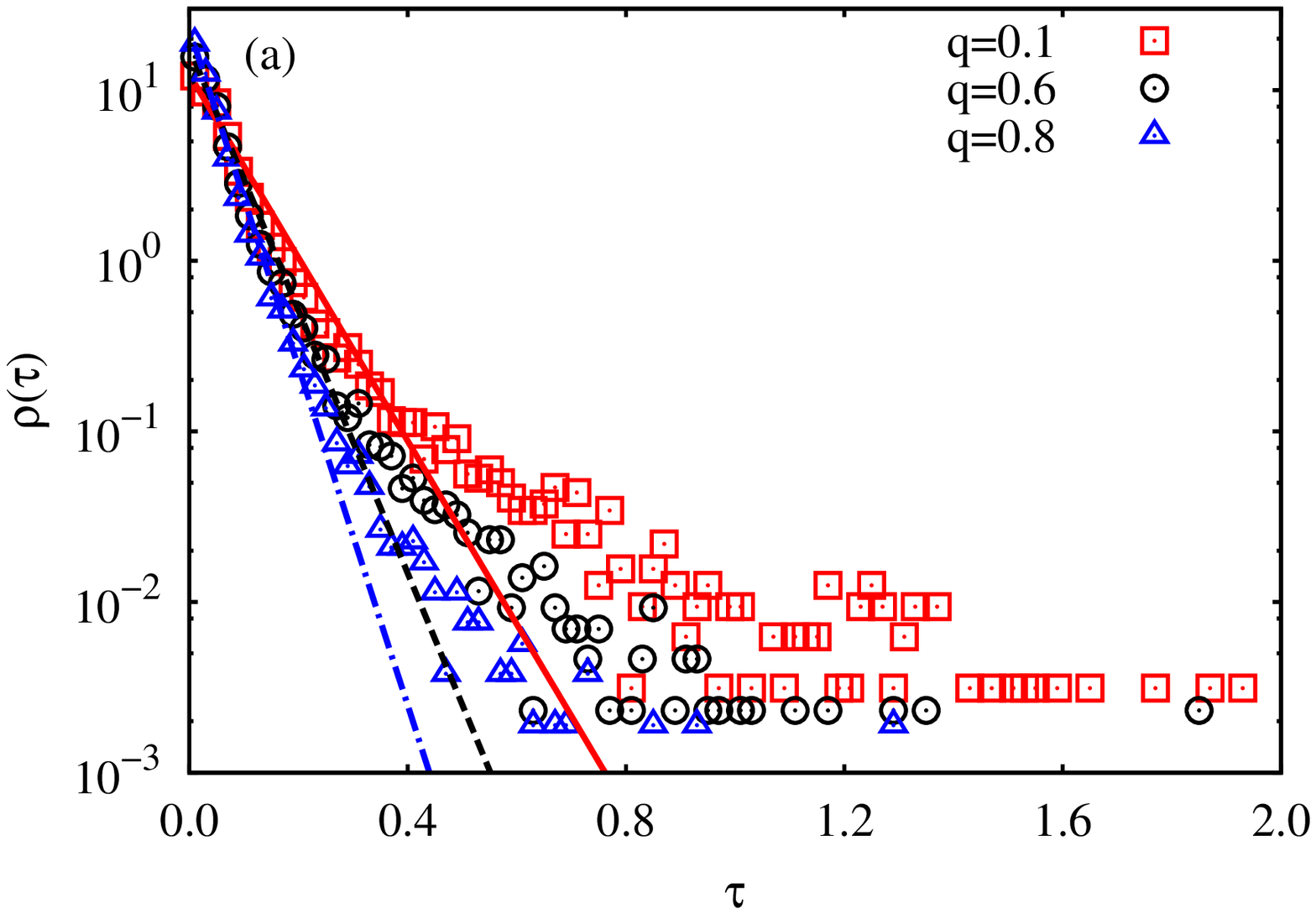}
\includegraphics[scale=0.42]{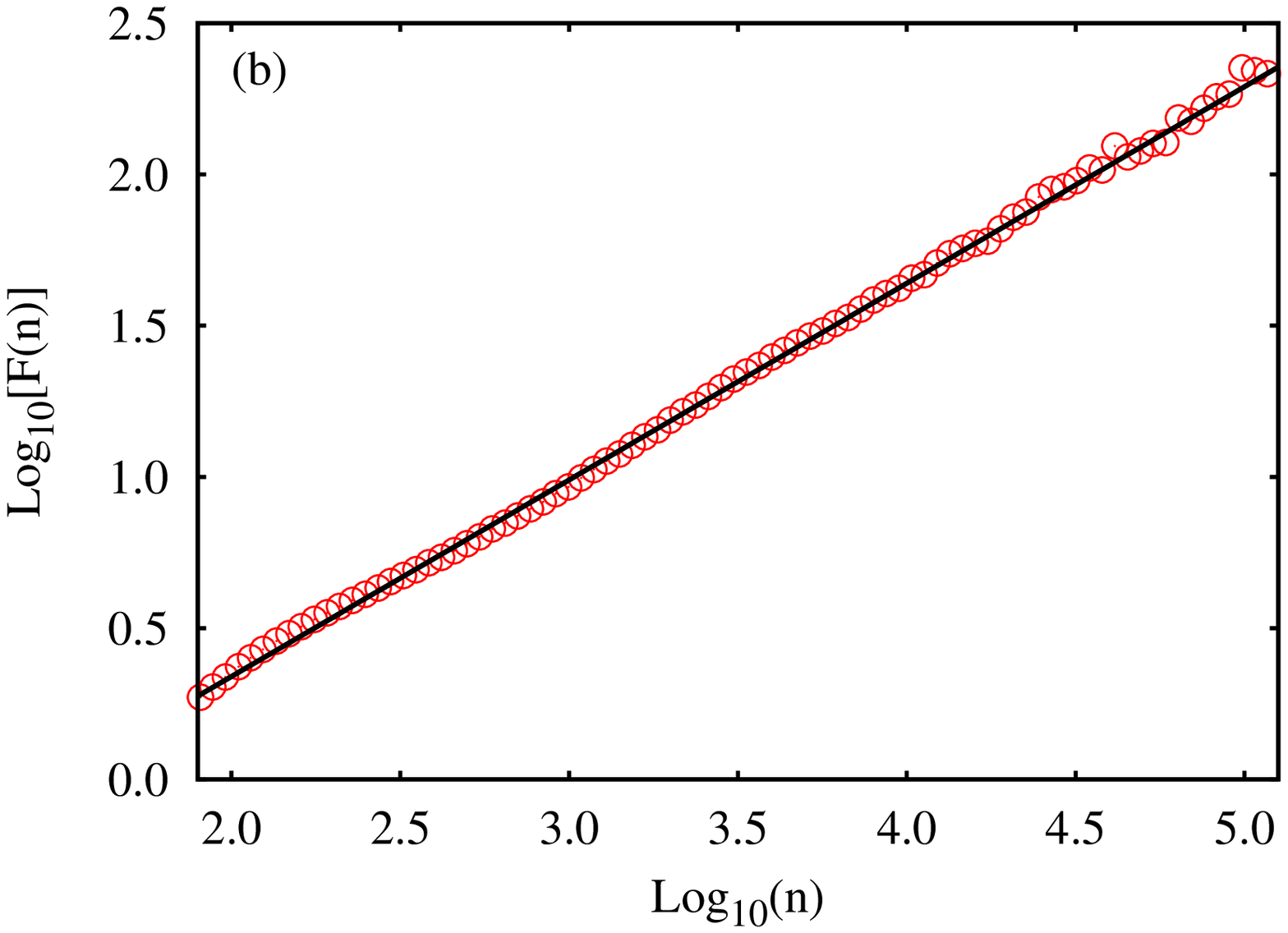}
\includegraphics[scale=0.42]{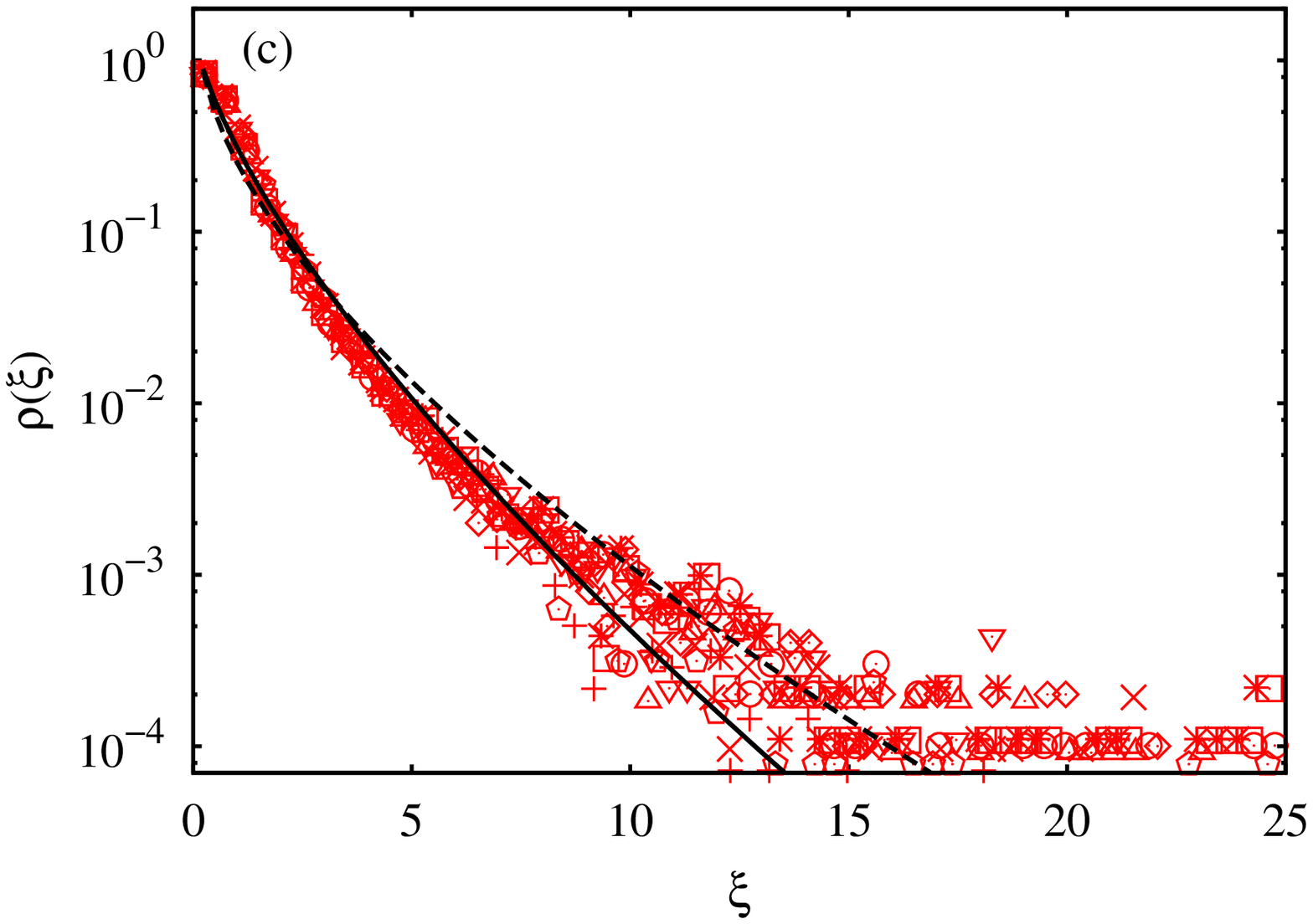}
\includegraphics[scale=0.42]{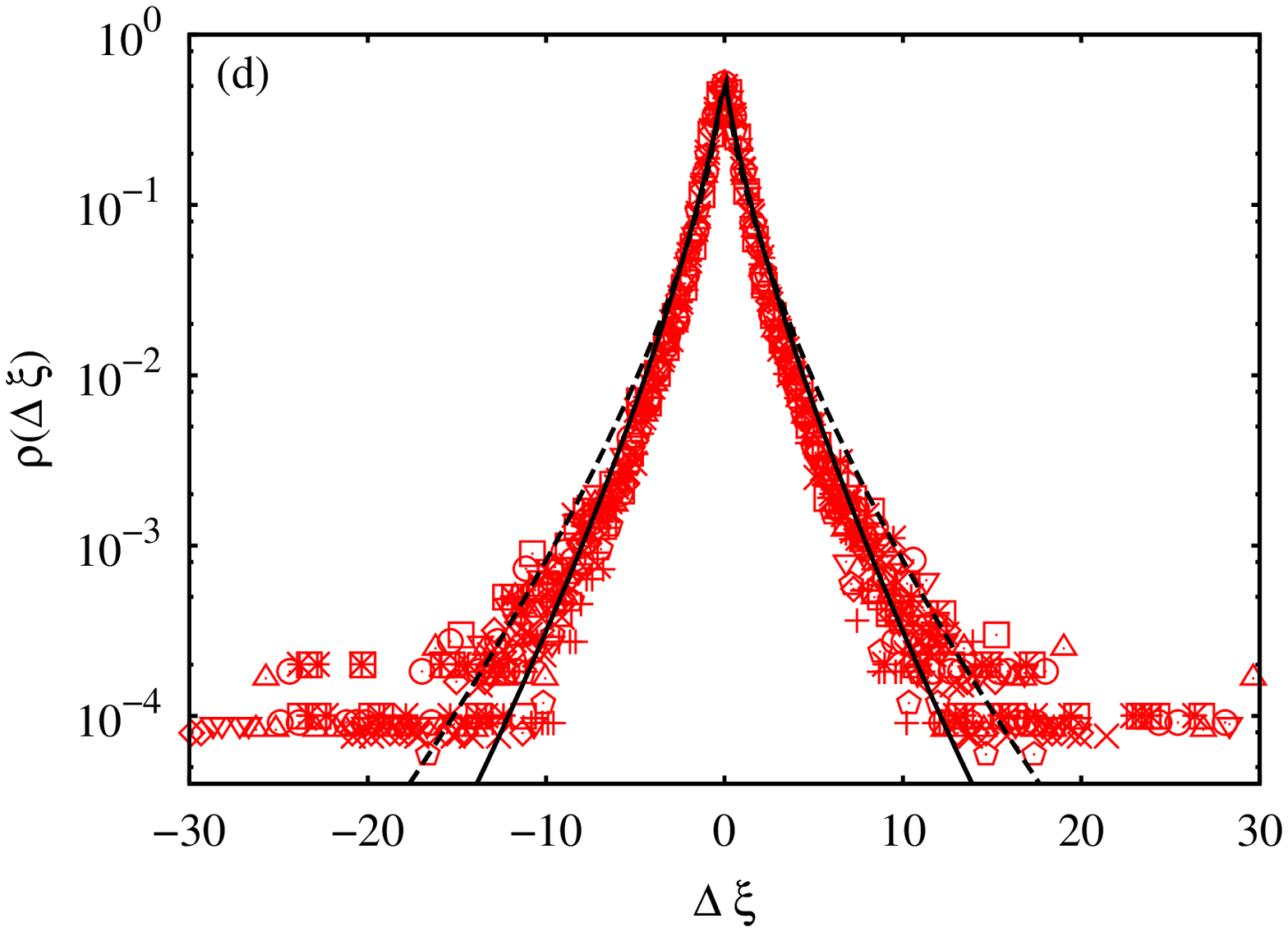}
\caption{(a) Probability distribution of the return interval $\tau$ for three values of $q$ (indicated in the figure)
in comparison with the exponential distribution of Eq.\ref{eq:exp}
for $\bar{\tau}_{0.1}=0.0808$ (continuos line), $\bar{\tau}_{0.6}=0.0564$ (dashed line) and 
$\bar{\tau}_{0.8}=0.0438$ (dashed-dotted line). 
(b) DFA analysis for the laser intensity signal: $\log_{10}[F(n)]$ versus $\log_{10}(n)$ in comparison with a linear fit, where we found  
$F(n)\propto n^h$ with $h\approx0.65$.
(c) Distribution of scaled variable $\xi=\tau / \bar{\tau}_q$ for eight equally spaced threshold values from 0.1 to 0.9
in comparison with a stretched exponential (continuos line) and with a Weilbull distribution 
(dashed line) both with $\gamma=2(1-h)=0.7$.
(d) Distribution of the return interval increments $\Delta \xi=\xi_{i+1}-\xi_{i}$ compared with Eq.\ref{cross} when considering the stretched exponential (continuos line)
and the Weilbull distribution (dashed line) both with $\gamma=0.7$. 
{ Notice that in the tail of the distributions the noise increases. This occurs because of the finite size of the set $\{\tau_i\}$
and also due to the small probability of finding large time intervals.}}
\label{fig:pdfs}
\end{figure*}

A possible manner to obtain these extreme events is considering a threshold value $q$ and archiving the initial times $t_i$ for which laser signal
is below this edge. 	The difference between two consecutive times, $\tau_i = t_{i+1}-t_{i}$, is the so called return interval.
This procedure is presented in Figure \ref{fig:exp}, where the right panel shows horizontal line segments representing the return intervals for $q=0.5$.
Figure \ref{fig:pdfs}a displays the probability density functions (pdf) of $\tau_i$, $\rho (\tau)$,  for three values of $q$. 
Clearly, the distribution is dependent on $q$, and it is also well known that for Gaussian random uncorrelated variables the return interval $\tau_i$  
is exponentially distributed according to\cite{Santhanam}
\begin{equation}\label{eq:exp}
\rho (\tau) = \frac{1}{\bar{\tau}_q}  e^{-{\tau}/{\bar{\tau}_q}}\,,
\end{equation}
where $\bar{\tau}_q$ is the mean value of return interval related to the  threshold value $q$. 
Figure \ref{fig:pdfs}a shows our data compared with this pdf, resulting in a poor agreement.

The weak agreement indicate that memory effects can be present in the bubble dynamics. To
address this question, we may use the detrended fluctuation analysis (DFA)\cite{Peng}. It basically 
consists of calculating the 
root mean square fluctuation function $F(n)$ (see for instance \cite{Kantelhardt}) for the integrated 
and detrended time series for different values of the time scale $n$. When we have scale-invariant
time series, $F(n)$ follows a power law behavior $F(n)\sim n^h$, where $h$ measures
the degree of correlation in the time series: if $h=0.5$, the series is uncorrelated,
while $h>0.5$ indicates long-range correlations. Figure \ref{fig:pdfs}b presents the results
concerning the laser signal where we found $h \approx 0.65$ leading to long-range correlations.

Empirical results have claimed that in the presence of power law correlation in the data the pdf $\rho (\tau)$ is 
usually adjusted by a stretched exponential\cite{Bunde,Yamasaki,Wang} or by  a Weilbull distribution\cite{Blender}, i.e.,
\begin{equation}\label{eq:stretched}
\rho (\tau) \sim e^{-A \left({\tau}/{\bar{\tau}_q}\right) ^\gamma} ~~\text{or}~~ \rho (\tau) \sim 
\left({\tau}/{\bar{\tau}_q}\right) ^{\gamma-1}e^{-B \left({\tau}/{\bar{\tau}_q}\right) ^\gamma}\,,
\end{equation}
where $A$ and $B$ are constants and $\gamma$ is the exponent of the power law autocorrelation function. 
These two distributions also emerge in the analytical approach of  Santhanam and Kantz\cite{Santhanam} when
considering a Gaussian fractional noise with autocorrelation exponent  $\gamma$. 
It is interesting to note that by employing the scaled variable $\xi=\tau / \bar{\tau}_q$ both distributions become independent on $q$
and that the exponents $\gamma$ and $h$ are related via $\gamma=2(1-h)$.
Figure \ref{fig:pdfs}c displays the distributions of the scaled variable $\xi$ and also the stretched exponential and the Weilbull distributions, which, due to the normalization
and the unitary mean value of $\xi$, have just one parameter, $\gamma$, determined from $h$. 
From this figure, we observe a good data collapse but a poor agreement with both previous distributions, specially for small $\xi$. In this case,
the distributions of Eq.\ref{eq:stretched} underestimate the data for $\xi\in [0,2]$ and overestimate for $\xi\in [3,7]$. The agreement is not improved if we find 
$\gamma$ via least square method. 

Another aspect to be investigated is the scaled return interval increments $\Delta \xi=\xi_{i+1}-\xi_{i}$. We know that the
pdf of the difference between two independent random numbers $X-Y$ is given by the cross-correlation\cite{Rohatgi}
\begin{equation}\label{cross}
f_{X-Y}(\tau)=\int_{-\infty}^{\infty} f_X(x) f_Y(x+\tau) dx\,, 
\end{equation} 
where $X$ is distributed according to $f_X(x)$ and $Y$ according to $f_Y(x)$. In the case of $\Delta \xi$, both
distributions are the same and we can use a stretched exponential or the Weibull of Eq.\ref{eq:stretched} to compare with the data.
Figure \ref{fig:pdfs}d shows this comparison, again finding an imperfect agreement. In addition, we have to mention that
the return interval series is week correlated ($h\approx 0.55$), thus the previous equation should be viewed as an approximation.
 
\section{Modeling}
In principle, we should be able to describe the dynamics of boiling fluid since the physical transport phenomena are well known to
follow the Navier-Stokes equation. However, technical difficulties as numerical instability because of the complex boundary of the two phases and
the large amount of simulation time required to do the integrations make this task very difficult. Thus, our goal is to understand this complex
phenomenon from a minimalist model. Therefore, we retain only the relevant ingredients to reproduce the main aspects of the experimental data.

In an approximate scenery, we can employ a two-state model for which 
only matters whether there are bubbles in the laser path or not. When
there are bubbles, the laser bean is considered totally scattered and the passing intensity 
signal is zero. On the other hand, when there are no bubbles, the
beam passes without scattering and the intensity signal is one.

In addition to the above two-state approximation for the laser signal, a significant ingredient of these empirical data is the power law
correlation presented in Figure \ref{fig:pdfs}b. It has been reported that a symbol series (here 0 or 1)  can present long-range correlations when it is generated by using only two uncorrelated random numbers\cite{Buiatti,Ribeiro}. In this context,  a possible way to model our data is considering a one-dimensional 
lattice where the sites represent the laser intensity signal in a given time. In the first
site of the lattice we start to fill it drawing a discrete random number from a Bernoulli distribution with probability parameter $p$. If the random 
number is 1, there are bubbles in the laser patch and the laser signal is zero for this time. Otherwise, there are no bubbles in the laser path and the laser
signal is one for this time and also for the next $[x]$ times (sites), where $[x]$ is the integer part of a random number $x$ distributed according to a distribution $P(x)$. 
All of the other sites
of this lattice are filled by repeating the above procedure. Due to the fact that the  simulated laser signal is only zeros or ones, it does not depend on the threshold 
value $q$. Actually, in this approach the return interval $\tau$ is exactly the length of the consecutive zero sites. Thus, we effectively focus our attention
on the the scaled variable $\xi$.

\begin{figure*}[!ht]
\centering
\includegraphics[scale=0.42]{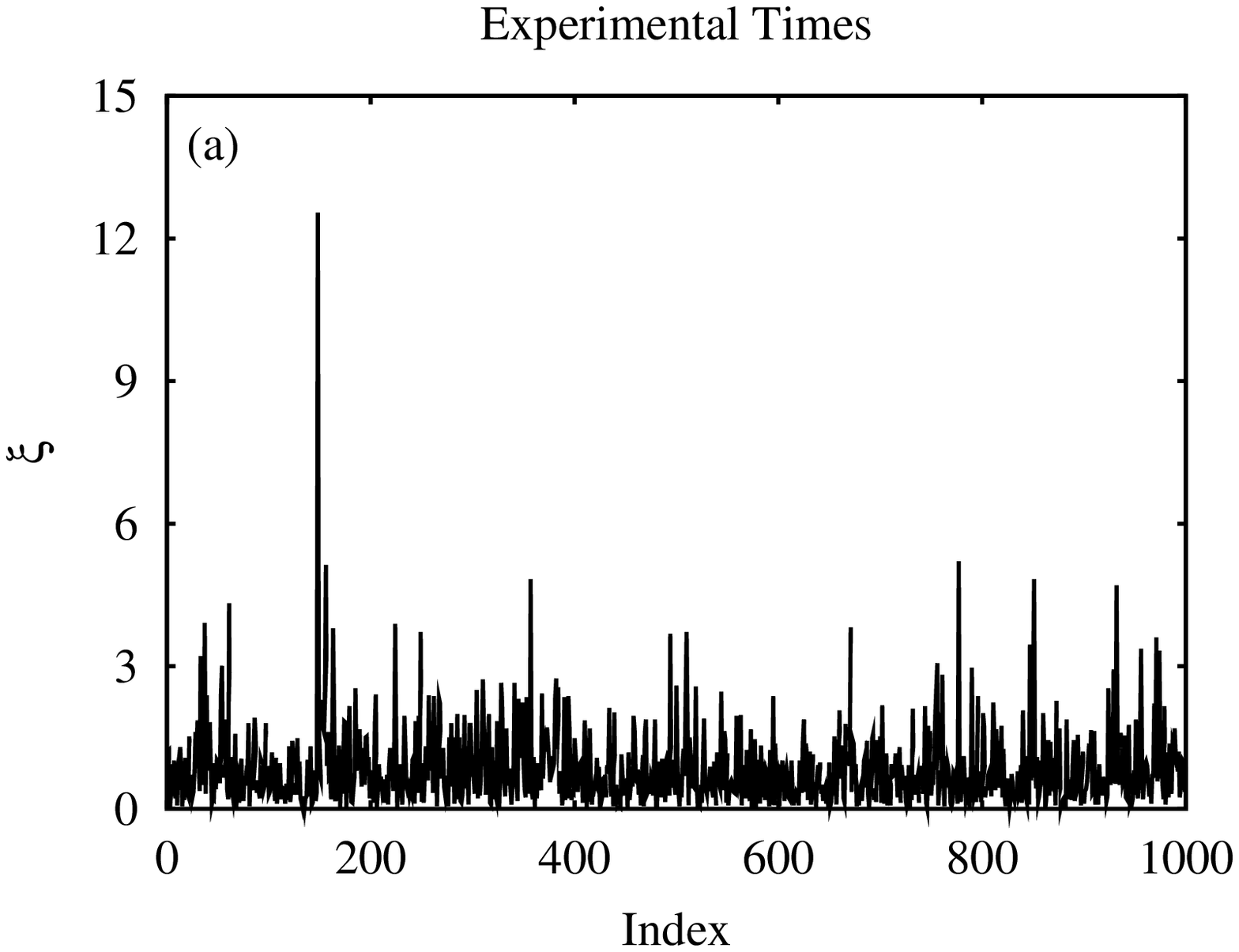}
\includegraphics[scale=0.42]{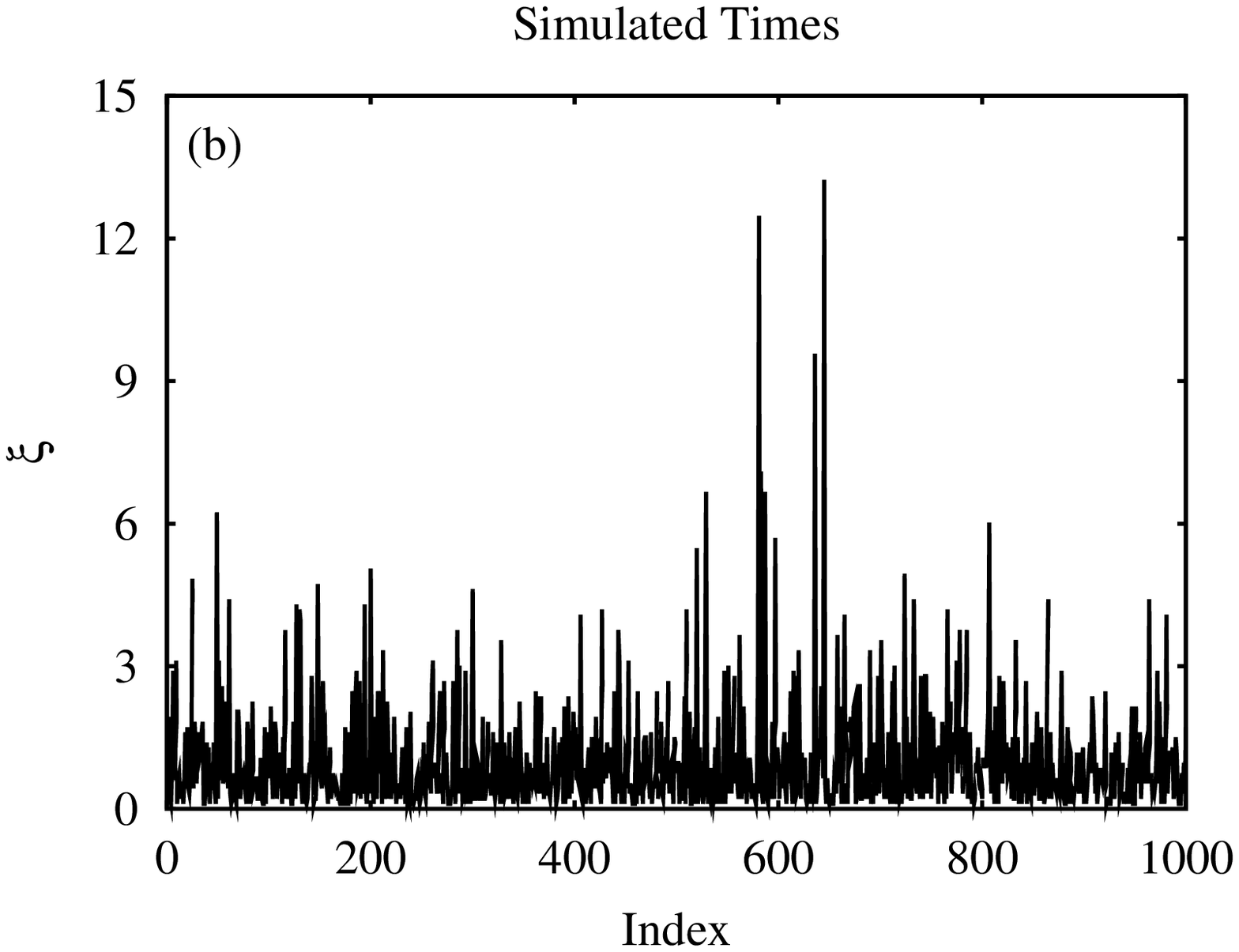}
\includegraphics[scale=0.42]{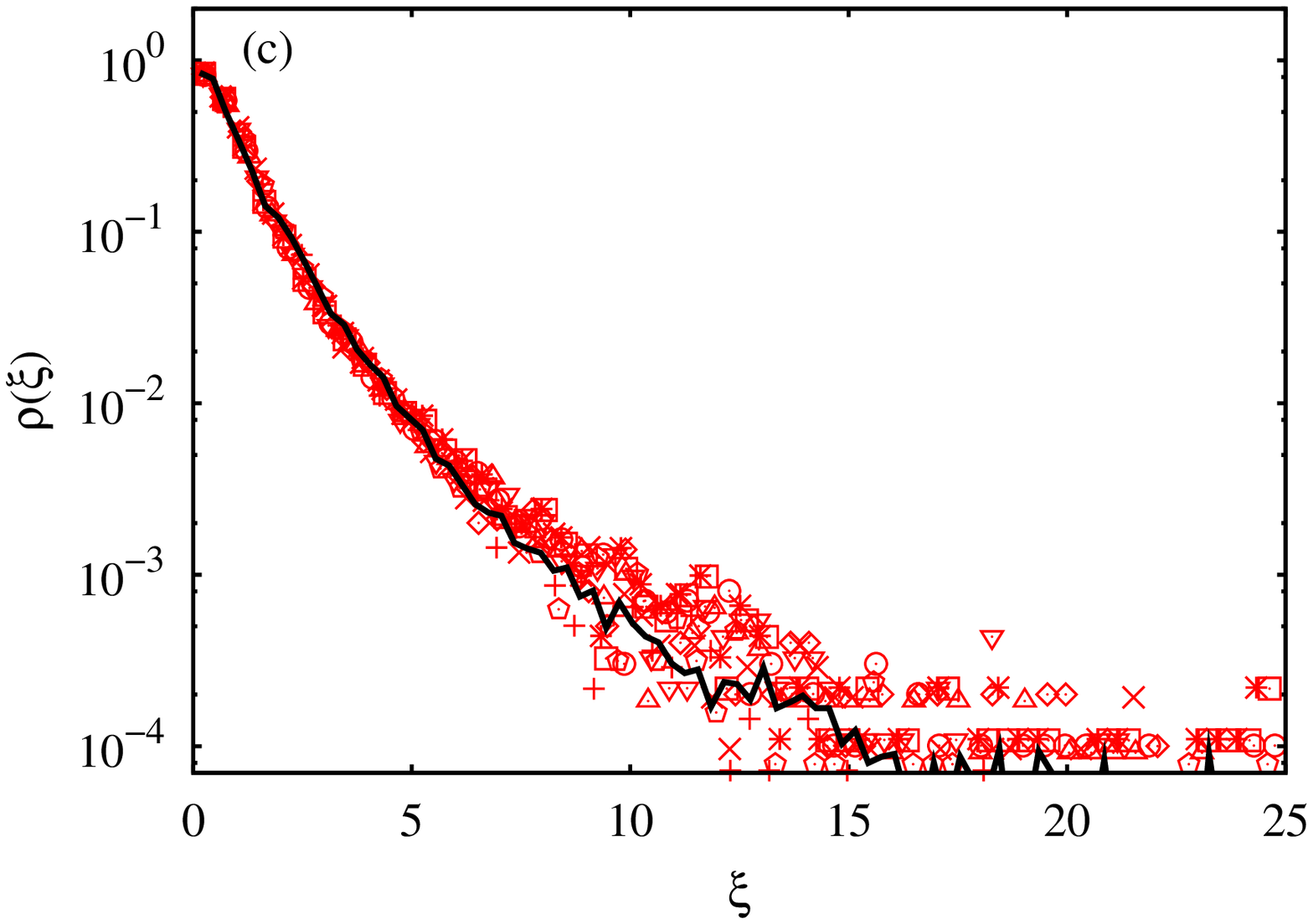}
\includegraphics[scale=0.42]{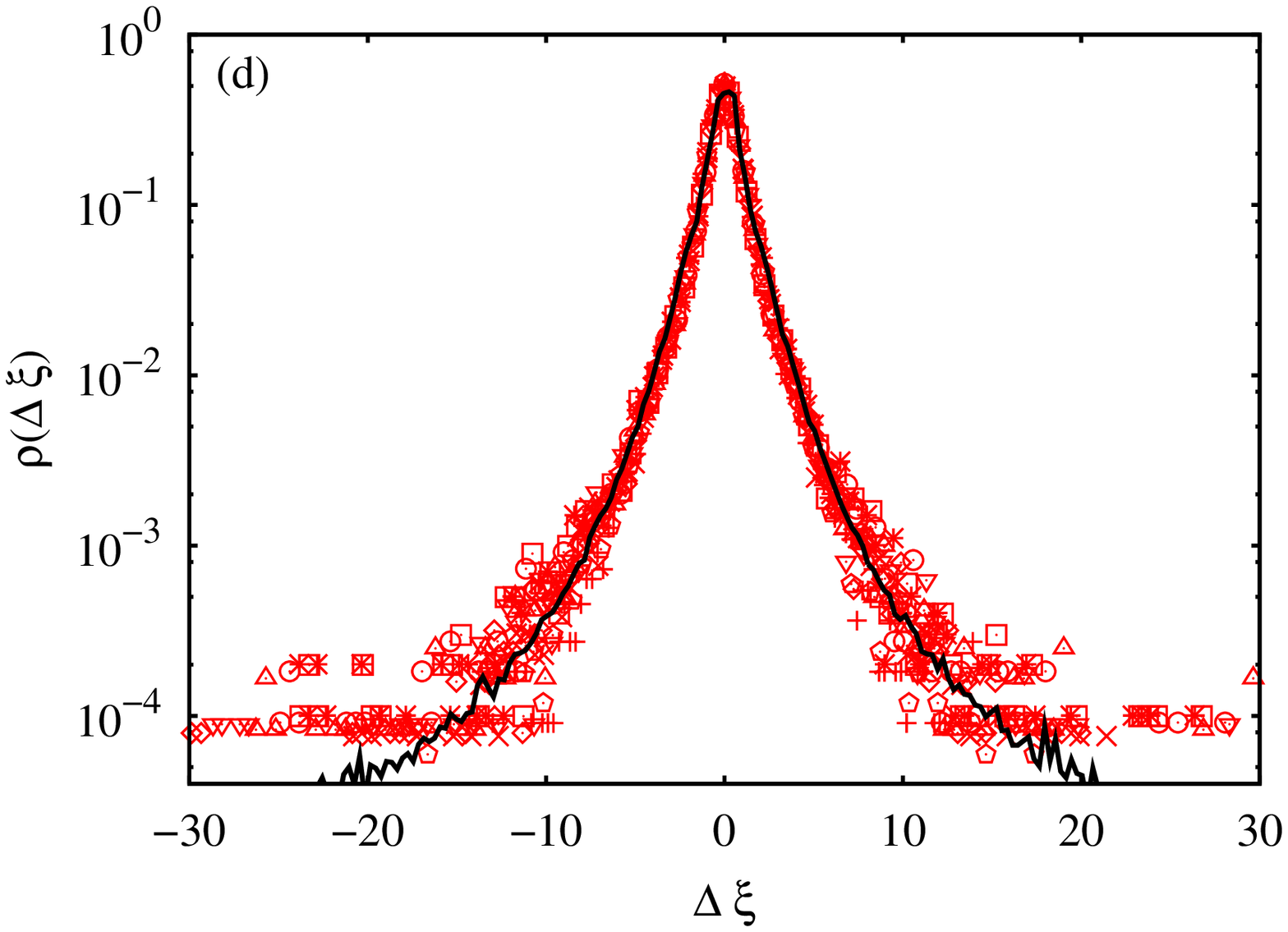}
\caption{A comparison between the experimental (a) and the simulated (b) return intervals. 
The simulated result was obtained by considering a Pareto distribution with $k=1$ and $\alpha=1.8$ and
the parameter $p=0.2$ for a lattice with $5 \times 10^4$ sites averaged over $200$ realizations. Figure (c) shows
the distribution of the return interval for eight equally spaced threshold values from 0.1 to 0.9 compared with
the simulated result (continuos lines). Figure (d) presents the comparison for the return interval increments $\Delta \xi=\xi_{i+1}-\xi_{i}$.}
\label{fig:comparison}
\end{figure*}

The proposed model appears to be very \textit{ad hoc} since we have a function to choose plus a parameter. However, when performing the simulations,
we empirically found that short tail distributions such as exponential, Gaussian, lognormal, and gamma are not able to improve the agreement found
when comparing with the analytical expressions of Eq.\ref{eq:stretched}. In contrast, when considering the most simple distribution with
power law tail, i.e., the Pareto one
\begin{equation}
P(x) = 
\begin{cases}
\alpha  k^{\alpha } x^{-\alpha -1}  & \text{if $x>k$,} \\
0 & \text{if $x<k$,}
\end{cases}
\end{equation}
where $k>0$, $\alpha>0$ are parameters of the distribution, the agreement with our data is very good.

Figure \ref{fig:comparison}
shows the simulated return interval in comparison with experimental data.
For the simulation, we have fixed the value $k=1$ and the size of lattice in a such way to have $5 \times 10^4$ return intervals, 
typically the number found in the experimental data. To obtain the best fit parameters, we incrementally update the values of $\alpha$ 
and $p$ getting the distribution $\rho(\xi)$ which is averaged over $200$ realizations and so confronted with the experimental data 
via the method of least squares. The best values found for these parameters are $\alpha=1.8$ and $p=0.2$. 

The previous model claims for an analytical approach. Indeed, it may be view as a sum of random numbers in which the number terms
is also a random number. Thus, supposing known the distribution of Pareto sums, we may write
\begin{equation}
\rho(\tau) = N \sum_{n=1}^{\infty} (1-p)^n S_n(\alpha_n,k_n)\,,
\end{equation}
where $N$ is the normalization factor, $(1-p)^n$ is the probability of summing $n$ consecutive numbers and $S_n(\alpha_n,k_n)$ is the 
distribution of Pareto sums.  However, to obtain a general expression for $S_n(\alpha_n,k_n)$ may not be a easy task\cite{Kluppelberg,Zaliapin,Nadarajah}.
It can be argued that these cumbersome calculations are avoided if we consider a stable distribution for $P(x)$. In fact, for this case the sum of stable
variables is well known to be a stable distribution, but the support of $P(x)$ is only positive when the stable index, $\alpha'$, is less that one and the
skewness parameter is equals to one\cite{Feller}. The stable distribution is asymptotically a power law with $P(x)\sim x^{-1-\alpha'}$ which is not compatible the exponent found when considering the Pareto distribution.

{ The comparison with the previous model also indicates that the laser signal is very close to a point possess,
since the deflection times are very short when compared with inter-event times $\tau_i$. By comparing
our model with Refs.\cite{Grigolini,Allegrini,Lowen}, we find that the laser signal can
be viewed as non-Poisson renewal process. This process is characterized by a sequence of events
spaced by time intervals that are independent random variables. Moreover, the time intervals
are draw from the same probability density $\psi(\tau)\sim\tau^{-\mu}$, where $\mu$ is the
renew index. Confronting this expression with the Pareto distribution, we can see
that the renew index is $\mu=\alpha +1\approx2.8$ for our case. In addition, following Grigolini \textit{et. al}\cite{Grigolini},
it is also possible to show that a fully asymmetric L\'evy stable distribution of index $\delta=(\mu-1)^{-1}\approx0.56$
emerges for number of events in an time interval $\tau'$. Further, the renew process may be also connected 
with a noise whose power spectrum is $1/f^\eta$, where $\eta=3-\mu\approx0.2$ in our case. 

The previous findings suggest that a $1/f^{0.2}$ noise is present in the context of a first-order phase transition, 
in contrast with the common association among inverse power laws noise and critical phenomena\cite{Chialvo,Mora}.
This has been also reported in the context of  first-order electronic phase transitions\cite{Ward} and for polymer
folding\cite{Chakrabarty}. From a general point of view, all the empirical discoveries suggest that, at boiling temperature,
the fluctuations between the two phases (here, the bubble and the liquid bulk) play an essential role in the system dynamics,
generating the nontrivial aspects reported here.
}

\section{Summary}
We have reported statistical analysis from the bubbles dynamics in a sample of boiling water.
Our analysis was focused on an experiment of laser scattering in which a laser beam
passed through the boiling fluid having its intensity monitored. By using this time series we
evaluated the return interval distribution finding a non-exponential distribution. In addition,
we verified that the dynamical processes that generates the bubbles introduces nontrivial correlations.
Employing a minimalist phenomenological model we were able to reproduce the experimental
behavior successfully . The model also seems to suggest  that a fundamental ingredient generating
the nontrivial dynamics is the power law tail related to the waiting time for the bubbles passing through the
laser path and also the correlations introduced by the draw of the two random numbers.

\section*{Acknowledgements}
We thank CNPq/CAPES for financial support and CENAPAD-SP for computational support.


\begin{thebibliography}{99}
\bibitem{Prosperetti} Prosperetti A.  Bubbles. Phys. Fluids 2004;16:1852-1865.
\bibitem{Lohse} Lohse D. Bubble puzzles. Physics Today 2003;56:36-41.
\bibitem{Joshi} Joshi JB. Computational flow modelling and design of bubble column reactors. Chem. Eng. Sci. 2001;56:5893-5933.
\bibitem{Alvarez} Alvarez-Ramirez J, Espinosa-Paredes G, Vazquez A. Detrended fluctuation analysis of the neutronic power from a nuclear reactor.  Physica A 2005;351:227-240
\bibitem{Brenner} Brenner MP, Hilgenfeldt S, Lohse D. Single-bubble sonoluminescence. 2002;74:425-484.
\bibitem{Seychelles} Seychelles F, Amarouchene Y, Bessafi M, Kellay H. Thermal convection and emergence of isolated vortices in soap bubbles. Phys. Rev. Lett. 2008;100. 144501,1-4
\bibitem{Vandewalle} Vandewalle N, Lentz JF, Dorbolo S, Brisbois F. Avalanches of popping bubbles in collapsing foams. Phys. Rev. Lett. 2001;86:179-182.
\bibitem{Ritacco} Ritacco H, Kiefer F, Langevin D. Lifetime of bubble rafts: Cooperativity and avalanches. Phys. Rev. Lett. 2007;98. 244501,1-4.
\bibitem{Schmidt} Schmidt LE, Keim NC, Zhand WW, Nagel SR. Memory-encoding vibrations in a disconnecting air bubble.  Nature Phys. 2009;5:343-346.
\bibitem{Tryggvason} Tryggvason G, Esmaeeli A, Lu J, Biswas S. Direct numerical simulations of gas/liquid multiphase flows. Fluid Dynam. Res. 2006;38:660-681.
\bibitem{Zahn} Zahn D. How does water boil? Phys. Rev. Lett. 2004;93. 227801,1-4.
\bibitem{BrennerC} Brennen CE. Cavitation and Bubble Dynamics. New York: Oxford University Press; 1995.
\bibitem{Shoji} Shoji M. Studies of boiling chaos: a review. Int. J. Heat Mass Transfer 2004;47:1105-1128.
\bibitem{Thome} Thome JR. Boiling in microchannels: a review of experiment and theory. Int. J. Heat Fluid Flow 2004;25:128-139.
\bibitem{Cordonet} Cordonet A, Lima R, Ramos E. Two models for the dynamics of boiling in a short capillary tube. Chaos 2001;11:344-350.
\bibitem{Iida} Iida Y, Lee J, Kozuka T, Yasui K, Towata A, Tuziuti T. Optical cavitation probe using light scattering from bubble clouds. Ultrason. Sonochem. 2009;16:519-524.
\bibitem{Gumbel} Gumbel EJ. Statistics of Extremes. New York: Dover Publications Inc.; 2004.
\bibitem{Galambos} Galambos J. The Asymptotic Theory of Extreme Order Statistics. New York: John Wiley \& Sons Inc; 1978.
\bibitem{Reiss} Reiss RD, Thomas M, Reiss RD. Statistical Analysis of Extreme Values: From Insurance, Finance, Hydrology and Other Fields. Boston: Birkhauser; 1997.
\bibitem{Embrechts} Embrechts P, Kluppelberg C, Mikosch T. Modelling Extremal Events for Insurance and Finance. New York: Springer; 1997.
\bibitem{Bunde} Bunde A, Eichner JF, Kantelhardt JW, Havlin S. Long-term memory: A natural mechanism for the clustering of extreme events and anomalous residual times in climate records. Phys. Rev. Lett. 2005;94. 048701,1-4.
\bibitem{Yamasaki} Yamasaki K, Muchnik L, Havlin S, Bunde A, Stanley HE. Scaling and memory in volatility return intervals in financial markets. Proc. Natl. Acad. Sci. U.S.A. 2005;102:9424-9428.
\bibitem{Wang} Wang F, Yamasaki K, Havlin S, Stanley HE. Scaling and memory of intraday volatility return intervals in stock markets. Phys. Rev. E 2006;73. 026117,1-8.
\bibitem{Blender} Blender R, Fraedrich K, Sienz F. Extreme event return times in long-term memory processes near 1/f. Nonlinear Processes Geophys. 2008;15:557-565.
\bibitem{Santhanam} Santhanam MS, Kantz H. Return interval distribution of extreme events and long-term memory. Phys. Rev. E 2008;78. 051113,1-9.
\bibitem{Peng} Peng CK, Buldyrev SV, Havlin S, Simons M, Stanley HE, Goldberger AL. Mosaic organization of DNA nucleotides. Phys. Rev. E 1994;49:1685-1689.
\bibitem{Kantelhardt} Kantelhardt JW, Koscielny-Bunde E, Rego HHA, Havlin S, Bunde A. Detecting long-range correlations with detrended fluctuation analysis. Physica A 2001;295:441-454.
\bibitem{Rohatgi} Rohatgi VK. Statistical Inference. New York: John Wiley; 1984.
\bibitem{Buiatti} Buiatti M, Grigolini P, Palatella L. A non extensive approach to the entropy of symbolic sequences. Physica A 1999;268:214-224.
\bibitem{Ribeiro} Ribeiro HV, Lenzi EK, Mendes RS, Mendes GA, da Silva LR. Symbolic sequences and Tsallis entropy. Braz. J. Phys. 2009;39:444-447.
\bibitem{Kluppelberg} Kluppelberg C, Mikosch T. Large deviations of heavy-tailed random sums with applications in insurance and finance. J. Appl. Probab. 1997;34:293-308.
\bibitem{Zaliapin} Zaliapin IV, Kagan YY, Schoenberg FP. Approximating the distribution of Pareto sums. Pure Appl. Geophys. 2005;162:1187-1228.
\bibitem{Nadarajah} Nadarajah S, Ali MM. Pareto random variables for hydrological modeling. Water Resour. Manag. 2008;22:1381-1393.
\bibitem{Feller} Feller W. An Introduction to Probability Theory and Its Applications. Vol. II.  New York: John Wiley; 1966.
\bibitem{Grigolini} Grigolini P, Palatella L, Raffaelli G. Asymmetric anomalous diffusion: an efficient way to detect memory in time series.
Fractal 2001;9:439-449.
\bibitem{Allegrini} Allegrini P, Menicucci D, Bedini R, Fronzoni L, Gemignani A, Grigolini P, West BJ, Paradisi P. Spontaneous brain activity as a source of ideal $1/f$ noise. Phys. Rev. E. 2009;80. 061914,1-13.
\bibitem{Lowen} Lowen SB, Teich MC. Fractal renewal processes generate $1/f$ noise. Phys. Rev. E 1993;47:992-1001.
\bibitem{Chialvo} Chialvo DR. Emergent complex neural dynamics. Nature Phys. 2010;6:744-750.
\bibitem{Mora} Mora T, Bialek W. Are biological systems poised at criticality? arXiv:1012.2242v1 2010.
\bibitem{Ward} Ward TZ, Zhang XG, Yin LF, Zhang XQ, Liu M, Snijders PC, Jesse S, Plummer EW, Cheng ZH, Dagotto E,  Shen J.
Time-resolved electronic phase transitions in manganites. Phys. Rev. Lett. 2009;102. 087201,1-4.
\bibitem{Chakrabarty} Chakrabarty S, Bagchi B. Temperature dependent free energy surface of polymer folding from
equilibrium and quench studies. J. Chem. Phys. 2010; 133. 214901,1-8.
\end{thebibliography}
\end{document}